\documentclass[twocolumn,showpacs,prl,10pt]{revtex4}
\usepackage[english]{babel}
\usepackage{amsmath,amssymb}
\usepackage{times}
\usepackage{epsfig}

\begin{document}

\title {Magnetic response  of nonmagnetic impurities in cuprates}
\author{P. Prelov\v sek$^{1,2}$ and I. Sega$^1$}
\affiliation{$^1$J.\ Stefan Institute, SI-1000 Ljubljana,
Slovenia}
\affiliation{$^2$Faculty of Mathematics and Physics, University
of Ljubljana, SI-1000 Ljubljana, Slovenia} 
\date{\today}

\begin{abstract}
A theory of the local magnetic response of a nonmagnetic impurity in a
doped antiferromagnet, as relevant to the normal state in cuprates, is
presented. It is based on the assumption of the overdamped collective mode
in the bulk system and on the evidence, that equal-time spin correlations
are only  weakly renormalized in the vicinity of the impurity. The theory
relates the Kondo-like behavior of the local susceptibility to the
anomalous temperature dependence of the bulk magnetic susceptibility,
where the observed increase of the Kondo temperature with doping
reflects the crossover to the Fermi liquid regime  and the
spatial distribution of the magnetization is given by bulk
antiferromagnetic correlations.

\end{abstract} 
\maketitle 

One of the open theoretical questions in cuprates is the understanding of a
well established experimental fact that nonmagnetic impurities have a
strong effect on superconducting as well as on the normal-state
properties of cuprates \cite{allo}. Prominent example is the
substitution of Cu in CuO$_2$ planes by Zn$^{2+}$ which acts as a
spinless impurity. In contrast to the naive picture of a weak scatterer,
Zn$^{2+}$ impurity depresses superconductivity already at small
concentration and represents a strong scatterer for transport
properties. In this contribution we mainly address the  
magnetic effects of an impurity in the  non-superconducting
phase. NMR experiments show that Zn induces 
local magnetic moments on the nearest neighbor (n.n.) copper sites
\cite{maja}, as well as on more distant copper neighbors
\cite{juli,ouaz}. The observed local susceptibility
\cite{bobr,mend} is well accounted for by the Curie-Weiss form
$\chi_{\rm loc} \propto 1/(T+\Theta)$, whereby $\Theta$ is independent of
the impurity concentration but reveals a clear dependence on doping.
Namely, in the underdoped YBaCu$_3$O$_{6+x}$ (YBCO) the behavior is
nearly Curie-like, with $\Theta \sim 0$, whereas in the overdoped YBCO
$\Theta$ shows a strong increase with doping \cite{bobr1}.  In analogy
with the Kondo effect in metals $\Theta$ has been interpreted as a
relevant Kondo temperature. Analogous effects in doped cuprates have
been established for a Li$^{+1}$ impurity which is also nonmagnetic
\cite{bobr1}. Since it has a different valence the similarity
indicates that the additional hole is not trapped near the impurity.
By a multi-nuclei NMR imaging it has become also increasingly clear
that the spatial distribution of the magnetization around the impurity
reflects the antiferromagnetic (AFM) correlations of the bulk system
\cite{ouaz}.

From the point of theoretical description, it seems well established that a
nonmagnetic impurity as, e.g., Zn$^{2+}$ can be incorporated into a
microscopic electronic model of CuO$_2$ planes in cuprates by introducing,
e.g., an inert empty site into the $t$-$J$ model relevant to cuprates
\cite{poil}, or a local site with a very different local energy within the
Hubbard model \cite{bulu,maie}. Impurity-induced moments and a Curie-like
susceptibility have been established in disordered spin systems, e.g., spin
ladders \cite{sigr}, in the presence of the gap in the spin excitation
spectrum. An analogous treatment, assuming a
spin gap in underdoped cuprates, also leads to the appearance of a
Curie-type susceptibility of an unpaired spin \cite{khal}. Kondo effect
based on the $t$-$J$ model and on the spin-charge separated ground state has
been used also to explain strong transport scattering on the impurity site
\cite{naga}. The 2D Hubbard model with impurities has been analysed using a
renormalized random-phase-approximation (RPA) \cite{bulu} to describe the
Knight-shift data in Zn and Li substituted YBCO. Numerical studies confirmed
the existence of an induced Curie susceptibility at low doping \cite{maie},
whereas variational Monte Carlo simulation established equal-time AFM
correlations around the impurity \cite{lian}.

Nevertheless, the understanding of magnetic effects of such an
impurity is far from satisfactory. It is clear that in the normal
state of cuprates there is at most a pseudogap in underdoped regime,
whereas at higher doping spin excitations are gapless and the usual
arguments for a Curie-type susceptibility from unparied spins
\cite{khal,naga} become unfounded. There is also no explanation for
the origin and the onset of finite and large Kondo temperature $\Theta>0$
in the overdoped regime. Lacking is also the theoretical answer to
the evident question whether the local magnetic response around the
impurity is just the reflection of the bulk (as well anomalous)
magnetic response, as evidenced by recent experiments \cite{ouaz}. In
any case, the $T$-dependence of magnetic response outside the spin gap
regime has been so far treated only within the RPA \cite{bulu}, which is 
generally not satisfactory enough to account for anomalous bulk
properties.

To address the above issues we present a generalization of the theory of
spin response applied to explain anomalous bulk
magnetic response \cite{prel}.  The novel input is the
observation that equal-time spin correlations around the impurity are
to a large extent unrenormalized. The consequence is that the local
magnetic response is  an image of the anomalous staggered susceptibility in
a homogeneous system, which exhibits also the Curie-Weiss behavior.
Consequently, the onset of a finite Kondo scale $\Theta$ is related to
the doping-driven crossover of the bulk spin dynamics from a
non-Fermi-liquid  spin dynamics to a Fermi-liquid one
\cite{bonc}.

Let us start with an approach to the spin response of a homogeneous
doped AFM. The dynamical spin susceptibility $\tilde \chi_{\bf
q}(\omega)=-\langle \!\langle S^z_{\bf q};S^z_{\bf q} \rangle
\!\rangle_{\omega}$ can be expressed within the memory function
formalism as \cite{prel}
\begin{equation}
\tilde \chi_{\bf q}(\omega)=\frac{-\eta_{\bf q}}{\omega^2+\omega 
M_{\bf q}(\omega) - \omega^2_{\bf q}}\,, \label{chiq}
\end{equation}
where $\eta_{\bf q}=-{\dot\iota}\langle [S^z_{-\bf q}\,
,\dot{S}^z_{\bf q})]\rangle$, $\omega^2_{\bf q}=\eta_{\bf q}/\chi_{\bf
q} $ and $\chi_{\bf q}=\tilde \chi_{\bf q}(\omega=0)$ is the static
susceptibility. $\omega_{\bf q}$ represents the collective-mode
frequency in the case of low damping $\gamma_{\bf q}\sim
M^{\prime\prime}_{\bf q} (\omega_{\bf q}) <\omega_{\bf q}$. However,
in cuprates the collective mode is found to be overdamped throughout
the normal phase, i.e., $\gamma_{\bf q} > \omega_{\bf q}$.  Following
the evidence from the analysis of microscopic models, such as the
planar $t$-$J$ model \cite{sega,prel}, we assume also constant
$\eta_{\bf q}\sim \eta$ and $\gamma_{\bf q}\sim \gamma$ in the region
of interest near the AFM wavevector ${\bf q} \sim {\bf Q} =
(\pi,\pi)$. $\eta$ is closely related to energy, so it is quite
$T$-independent and only smoothly dependent on doping. In a doped AFM
damping $\gamma$ emerges from the decay of a spin collective mode into
electron-hole excitations and is also assumed to remain finite and
large at low $T \to 0$ in the normal state \cite{sega}. Hence, the
main $T$-variation enters Eq.~(\ref{chiq}) via $\chi_{\bf q}$.

The central idea of the theory of the anomalous scaling in doped AFM
\cite{prel} is that a nontrivial dependence $\chi_{\bf q}(T)$ is
driven by the fluctuation-dissipation relation for the equal-time
correlations
\begin{equation}
\frac{1}{\pi}\int_0^\infty d\omega ~{\rm cth}\frac{\omega}{2T}
\tilde \chi^{\prime\prime}_{\bf
q}(\omega)= \langle S^z_{-{\bf q}} S^z_{\bf q}\rangle = C_{\bf q}\, .
\label{eqsum}
\end{equation}
There exists an extensive evidence from analytical \cite{sing} and
numerical work on the $t$-$J$ model at finite hole concentration
$c_h>0$ that $C_{{\bf q} \sim {\bf Q}} \sim C/(\kappa^2+\tilde q^2)$,
$\tilde {\bf q}={\bf q}-{\bf Q}$, and the AFM correlation length
$\xi=1/\kappa$ saturates at low $T$. Similar
conclusions emerge from an analysis of inelastic neutron scattering on
YBCO \cite{bala,kaku}. The relation (\ref{eqsum}) then leads to a 
strongly $T$ dependent $\chi_{\bf q\sim\bf Q}(T)$ which is at heart of the 
non-Fermi liquid behavior observed in underdoped cuprates, in contrast
to the usual Fermi liquid where $\tilde \chi_{\bf
q}^{\prime\prime}(\omega)$ is essentially $T$-independent.

Let us first consider the behavior of $\chi_{\bf q}(T)$. Performing
the high-$T$ expansion in Eq.~(\ref{eqsum}) we get
\begin{equation}
C_{\bf q} \sim T \chi_{\bf q} + \frac{1}{\pi}\int_0^\infty d\omega
\frac{\omega}{6T} \tilde \chi^{\prime\prime}_{\bf q}(\omega)= T
\chi_{\bf q} + \frac{\eta_{\bf q}}{12T}, \label{ht}
\end{equation}
where we have taken into account the definition of $\eta_{\bf q}$ as
the second frequency moment of the shape function $\tilde
\chi^{\prime\prime}_{\bf q}(\omega)/\omega$. The high-$T$ expansion is
thus consistent with the Curie-Weiss behavior $\chi_{\bf Q}=C_{\bf
Q}/(T+\Theta_{\bf Q})$ where $\Theta_{\bf Q}=\eta_{\bf Q}/12 C_{\bf
Q}$. Since in a doped AFM, as represented, e.g., by the $t$-$J$ model,
$C_{\bf Q} \propto 1/c_h$ \cite{bonc} we get quite small scale
$\Theta_{\bf Q} \propto c_h$. On the other hand, the expansion is
valid only for $T>\Theta_{\bf Q}$. At $T=0$ we get $\chi_{\bf
Q}(T=0)=\eta/(\gamma \omega_p)$ \cite{prel}, where $\omega_p \sim
\gamma e^{-2 \zeta}$ and $\zeta = \pi \gamma C_{\bf Q}/2\eta$.

In Fig.~1 we present $C_{\bf Q}/\chi_{\bf Q}(T)$ as follows from the
solution of Eqs.~(\ref{chiq}),(\ref{eqsum}) for fixed $\eta=0.6~t$ and
$\gamma=0.5~t$ \cite{prel} (note that $t \sim 400~$meV for cuprates)
but varying $\kappa=0.5 - 1.5$ (in units of inverse lattice spacing). By way
of Eq.(\ref{eqsum}) $\kappa$ also determines 
$\tilde \kappa$ which enters the  low-$\omega$ behavior of
${\tilde\chi}^{\prime\prime}_{\bf q}(\omega)$. For the range of
$\kappa$ considered, $\kappa/\tilde\kappa\sim 2$ so that $\tilde\kappa$
roughly corresponds to the experimental range of values in underdoped to
overdoped YBCO \cite {bala}. 
Note that the Curie-Weiss law is  obeyed down to $T \sim
\Theta_{\bf Q}$. The deviation appears (to lower values for given
parameters) on approaching $T\to 0$. $\Theta_K=C_{\bf Q}/\chi_{\bf Q}(T=0)$
can be interpreted as the relevant Kondo 
temperature and its variation with $\kappa$ is presented in the inset
of Fig.~1. On increasing $\kappa$  (note that $\kappa \propto \sqrt{c_h}$)
we are facing quite an abrupt transition from a Curie behavior
at $T \to 0 $ to a Curie-Weiss variation with finite and rapidly
increasing $\Theta_K>0$ \cite{bonc}. Such a behavior is also
consistent with experimental results for $1/\chi_{\bf Q}(T)$ obtained
from the analysis of NMR $1/T_{2G}(T)$ relaxation in various cuprates
\cite{bonc}.

\begin{figure}[htb]
\centering
\epsfig{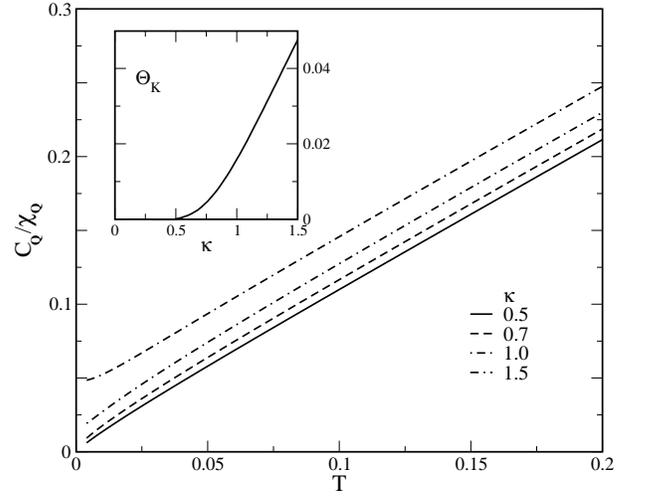}
\caption{$C_{\bf Q}/\chi_{\bf Q}$ vs. $T$ (both in units of $t$) for
various $\kappa$. The inset shows the variation of Kondo $\Theta_K$
with $\kappa$.}
\label{fig1}
\end{figure}

Let us turn to a doped AFM with an added nonmagnetic impurity, as
relevant to Zn$^{2+}$ replacing planar Cu in CuO$_2$ planes. To be
more specific we have in mind the planar $t$-$J$ model, where the impurity
is represented as an empty site at the origin $i=0$. In analogy with the
homogeneous system, Eq.(\ref{eqsum}), we consider the equal-time local
correlations $C_{ij}=\langle S_i^z S_j^z \rangle $ as an essential
input. In general, $C_{ij}$ differ from correlations in an homogeneous
system without any impurity where $C^0_{ij}=C^0({\bf R}_j-{\bf
R}_i)$. (Furtheron we lable the homogeneous quantities by the superscript
0.) However, it appears characteristic for 
strongly correlated electrons in low-doped AFM that at least shorter-range
$C_{ij}$ close to the impurity deviate modestly from unperturbed $C^0_{ij}$.
This is partly plausible since, e.g., the n.n. correlations $C_{ij}$ are
governed by the minimization of the exchange energy. On the other
hand, insensitivity of $C_{ij}$ around the impurity can be interpreted
as an effective spin-charge decoupling, where the empty site
represents just a free spinon not affecting nearby spin correlations
\cite{khal,naga}.

In support of the above conjecture we perform an
exact-diagonalization calculation of $C_{ij}$ within the $t$-$J$ model
with an empty site.  As an example of a doped AFM we present in Fig.~2
results for a system $N=20$ sites at $T=0$ with $N_h=2,3$ mobile holes
and $J/t=0.3$. We show some nonequivalent n.n. and next
n.n. correlations $C_{ij}$ around the impurity. We notice that even on
sites neighboring the impurity at least shorter-range $C_{ij}$ are
nearly the same as in the bulk, or even enhanced as noted also by
others \cite{lian}.

\begin{figure}[htb]
\centering
\epsfig{file=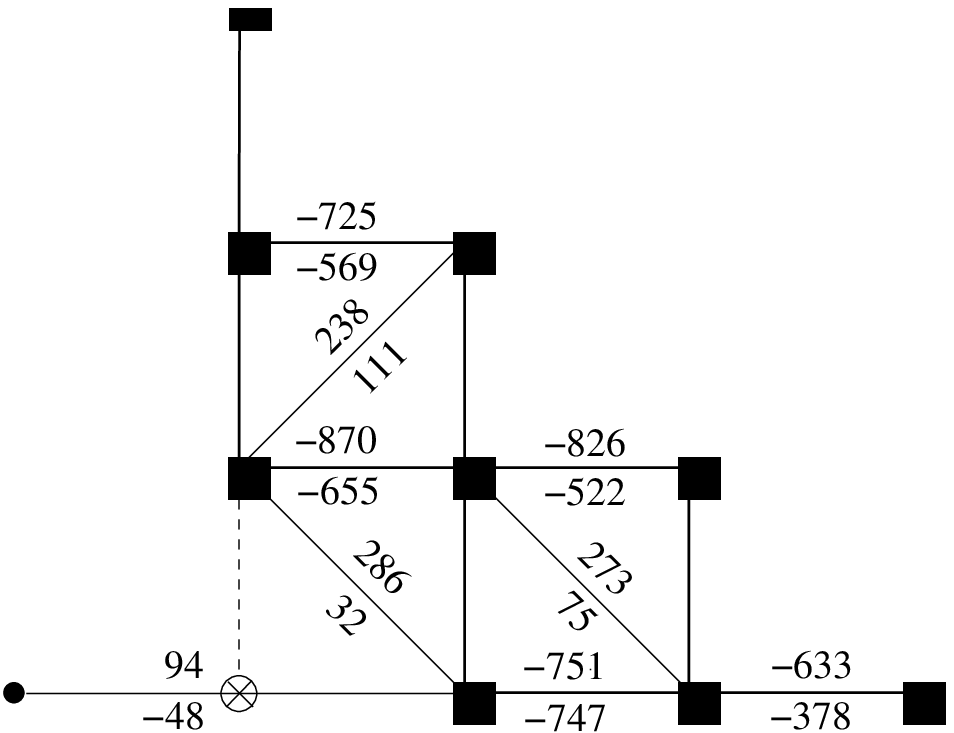,width=70mm}
\caption{Equal-time correlations $C_{ij}$ around the impurity (denoted
by $\otimes$) within the $t$-$J$ model for $J/t=0.3$ and dopings
$c_h=2/20$ (upper) $c_h=3/20$ (lower value) in units of $10^{-4}$.}
\label{fig2}
\end{figure}

It is therefore plausible to treat a system with an impurity by
assuming frozen correlations. I.e., we take that $C_{ij}=0$ for either
$i=0$ or $j=0$, while $C_{ij}=C^0_{ij}$ elsewhere. Note
that at $T=0$ and in a homogeneous system on a bipartite lattice (with
even $N,N_h$) the ground state is a singlet ($S_{\rm tot}=0$). In the
case of a single impurity this leads to
\begin{equation}
\sum_{i,j\neq 0} C_{ij}= \sum_{i,j} C^0_{ij}-2 \sum_{j} C^0_{ij}
+C^0_{00}= C^0_{00}=\frac{1}{4}-c_h, \label{cij}
\end{equation}
since $\sum_{j} C^0_{ij}= \langle S_i^z S_{\rm tot}^z \rangle = 0$.  
Eq.~(\ref{cij}) reproduces for $c_h \ll 1$ the proper moment of
the impurity corresponding to $S_{\rm tot}=1/2$ \cite{poil,khal}.

We  further study the real-space local susceptibilities
$\tilde \chi_{ij}(\omega)=-\langle\!\langle S^z_i; S^z_j
\rangle\!\rangle_{\omega}$ in analogy to the homogeneous case. The 
dynamical matrix $\underline{\tilde \chi}(\omega)$ can be generally expressed
in terms of corresponding matrices in real space $\underline
{\chi},\underline{\eta},\underline {M}$ as
\begin{equation} 
\underline {\chi}(\omega)=-[\omega^2 \underline 1 +\omega
\underline {M}(\omega) - \underline {\delta}]^{-1}\underline {\eta},
\label{chiloc}
\end{equation}
where $\underline{\delta}= \underline{\eta} \underline {\chi}^{-1}
$. Again, the local fluctuation-dissipation relation 
\begin{equation}
\frac{1}{\pi}\int_0^\infty d\omega ~{\rm cth}\frac{\omega}{2T} \tilde
\chi^{\prime\prime}_{ij}(\omega)= C_{ij}, \label{sumloc}
\end{equation}
is used to fix $\underline{\chi}$.

The next step in an inhomogeneous system is to diagonalize the
correlation matrix $C_{ij}=\sum_\lambda C_\lambda (v^i_\lambda)^*
v^j_\lambda$ where $\underline{C} {\bf v}_\lambda= C_\lambda {\bf
v}_\lambda$. From the sum rule (\ref{sumloc}) it then follows that one
can simultaneously diagonalize the susceptibility matrix $\tilde
\chi_{ij}(\omega)=\sum_\lambda \tilde \chi_\lambda(\omega) (v^i_\lambda)^*
v^j_\lambda$ with
\begin{equation} 
\tilde \chi_\lambda(\omega)=-\eta_\lambda/ (\omega^2+i \gamma_\lambda
\omega - \delta_\lambda), \label{chilam}
\end{equation}
where we have for simplicity assumed the diagonal form of 
$(M_{ij},\eta_{ij}) \sim \sum_\lambda (i\gamma_\lambda,\eta_\lambda)
(v^i_\lambda)^* v^j_\lambda$ as well. In analogy with the homogeneous system,
the local magnetic response around the impurity will be determined by
the behavior around $\lambda \sim \Lambda$ for which $C_\Lambda=~$max. 
In this region we assume constant 
$\gamma_\lambda \sim \gamma$ and $\eta_\lambda \sim \eta$. Clearly,
this is based on the assumption that  damping is quite local and not
affected significantly by the impurity. On the other hand,  $\eta_{ij}$ can
be explicitely calculated and is expressible in terms of
equal-time correlations \cite{sega}.

In a system with a nonmagnetic impurity a spin
polarization around the impurity induced by a homogeneous magnetic
field, gives rise to the susceptibility $\chi_i$ on site $i$
\begin{equation} 
\chi_i=\sum_j \chi_{ij}= \sum_{\lambda}\chi_\lambda
(v^i_\lambda)^* v_\lambda, \quad v_\lambda =
\sum_j v^j_\lambda. \label{chii}
\end{equation}
In the case of a n.n. site $\chi_i$ is directly related to the
Knight shift on the ${^7}$Li impurity and of $^{89}$Y near the Zn
impurity \cite{bulu}. Another relevant quantity is the (average)
uniform susceptibility $\bar \chi= \sum_\lambda \chi_\lambda
|v_\lambda|^2/N$ which yields the impurity-induced contribution $\Delta
\chi =\bar \chi-\chi_{q=0}^0$. 

We first give an approximate solution to the impurity problem via the
perturbation calculation. Here, $\underline C= \underline C^0 +
\underline C^\prime$ with the perturbatrive part $C_{00}^\prime =-1/4$
and $C_{0j}^\prime=C_{j0}^\prime = -C_{0j}^0$. The unperturbed
eigenvectors are the homogeneous ones, $v^i_{\bf q}=\exp(i{\bf q}{\bf
r}_i)/\sqrt{N}$, and the lowest order calculation gives
\begin{equation}
\Delta {\bf v}_{\bf q}= \frac{1}{N} \sum_{{\bf q}' \neq {\bf q}} {\bf
v}_{{\bf q}'} \frac{\mu^2-C^0_{\bf q}-C^0_{{\bf q}'} }
{ C^0_{\bf q}- C^0_{{\bf q}'}}, 
\label{pert}
\end{equation}
where $\mu=\sqrt{1/4-c_h}$ is the effective local moment. The
impurity-induced correction to the local susceptibility is thus
\begin{equation}
\Delta \chi_i= \sum_{\bf q} [\Delta \chi_{\bf q} v^i_{-{\bf q}}
v_{\bf q}+ \chi^0_{\bf q} (\Delta v^i_{-{\bf q}}
v_{\bf q}+ v^i_{-{\bf q}} \Delta v_{\bf q})],
\label{dchi}
\end{equation}
and can be expressed as $\Delta \chi_i=\Delta \chi/N+\tilde \chi_i$,
i.e., as a sum of the perturbed uniform susceptibility $\Delta \chi= N
\Delta C_{q=0,q=0}/T= \mu^2/T$, whereas
\begin{equation}
\tilde \chi_i= \sum_{\bf q} e^{i{\bf q}{\bf r}_i}
\chi^0_{\bf q} [\frac{\mu^2}{C^0_{\bf q} }-1].
\label{chil}
\end{equation}
It is evident that  $\Delta \chi$ corresponds to a free spin with the moment
$\mu \sim 1/2$ (at low doping) introduced by the impurity. Nontrivial is the
spatial distribution of $\tilde \chi_i$. Since the
main contribution in Eq.~(\ref{chil}) arises from ${\bf q} \sim
{\bf Q}$ and $C^0_{\bf Q} \gg \mu^2$, we obtain from Eq.~(\ref{chil}) $\tilde
\chi_i\sim -\chi^0_{0i}$, i.e., the local response is just the intersite
susceptibility of a homogeneous system, being the Fourier transform of
$\chi^0_{\bf q}$. 

In Fig.~3a we present results for $1/\tilde \chi_1$ for the n.n. site.
We use Eq.~(\ref{chil}) with $\chi_{\bf q}^0$ determined via a
self-consistent solution to Eqs.~(\ref{chiq}),(\ref{eqsum}).  To be
consistent with the $t$-$J$ model on a 2D square lattice we take here the
form $C^0_{\bf q}=a/(\kappa^2+\zeta_{\bf q})-b$ with $\zeta_{\bf
q}=2(\cos(qx) +\cos(qy)+2)$ whereas $a,b$ are chosen such that
$C^0_{q=0}=0$ and $C^0_{ii}=1/2$. As in the case of homogeneous
$\chi_{\bf Q}^0$ in Fig.~1, we notice that the behavior is close to the
Curie-Weiss form $1/\tilde \chi_1 \propto T+\Theta_1$ with a
qualitative transition from $\Theta_1 \sim 0$ for $\kappa
<0.7 $ to finite and large $\Theta_1$ for $\kappa>1$. Deviations from
the Curie-Weiss dependence are understandable since Eq.~(\ref{chil})
includes contributions from all ${\bf q}$ where $\Theta_{\bf q} >
\Theta_{\bf Q}$. Nevertheless the behavior at ${\bf q} \sim {\bf Q}$ is
dominant since considered $\kappa$ are quite large. It should be
reminded that the essence of the theory of anomalous scaling of spin
response in underdoped cuprates \cite{prel} lies in the fact that
anomalous $\omega,T$ dependence appears for $T>T_K \sim 0$ even for
substantial and $T$-independent $\kappa$.

\begin{figure}[htb]
\centering
\epsfig{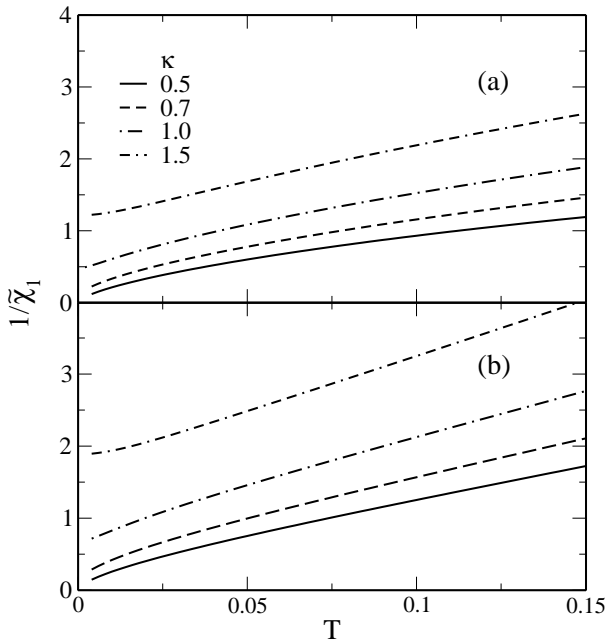}
\caption{Local susceptibility $1/{\tilde \chi}_1$ for the site n.n. to the
impurity vs. $T$ (both in units of $t$) for different $\kappa$: a) evaluated
via the perturbative calculation, b) with the diagonalization procedure. }
\label{fig3}
\end{figure}

Alternatively, we can solve the impurity problem as defined by
Eqs.(\ref{chiloc})-(\ref{chilam}) directly, without
relying on the perturbation expansion. We first find eigenvalues
$C_\lambda$ and corresponding ${\bf v}_\lambda$ numerically. For the
homogeneous doped system we assume that $C^0_{ij}$
correspond to 2D-lattice $C^0_{\bf q}$ as discussed above. The final results
for local ${\tilde \chi}_i$, as obtained from Eq.~(\ref{chii}) for the same
parameters as in the perturbation calculation, are shown in Fig.~3b. Note
that results presented in Figs.~3a,b are quite similar, even quantitatively.
However, the full calculation yields results even closer to the
Curie-Weiss behavior. 

Of interest is also the spatial distribution of the local
susceptibilities $\tilde \chi_i$ around the impurity. It is evident
from Eq.~(\ref{chil}) that it reflects the bulk
$\chi^0_{0i}$. Moreover, if we consider only the high-$T$ local Curie
constant ${\cal A}_i$, it follows from $\chi_{\bf q} \sim C^0_{\bf q}/T$
that ${\cal A}_i \sim - C^0_{0i}$. Thus ${\cal A}_i$ measures the equal-time
AFM correlations, changing sign in accordance with the sign of $C^0_{0i}$.
It is also plausible that the local response $\tilde 
\chi_i$ is quite sensitive to any deviations of
$C_{ij}-C^0_{ij}$ around the impurity. However, the latter affect mainly
the magnitude but not the $T$-dependence of $\tilde \chi_i$.
 
In summary, we have presented a theory for the magnetic response of
nonmagnetic impurities in doped AFM. It is based on the assumptions,
and also evidence, that certain quantities are not substantially
modified in the vicinity of an impurity. This is in particular the case
of correlations $C_{ij}$, but also of the  collective mode amplitude $\eta$
and damping $\gamma$. Note that such  assumptions would not be
valid within a normal Fermi liquid but are rather the consequence of
strong correlations, i.e., $C_{ij}$ exhibit a kind of charge-spin
separation.
 
Within the present theory the local spin response around the impurity
clearly reflects the one in the homogeneous system which is also anomalous. Or
alternatively, the measurements of the impurity-induced susceptibility
allow for the reconstruction of the bulk $\chi_{\bf q}$ and
corresponding correlation length $\xi$, as has been established in
some other systems with the spin gap \cite{tedo}.
 
We have also shown that $\chi_{\bf Q}$ in a doped AFM follows quite
well the Curie-Weiss behavior. Consequently also local ${\tilde \chi}_i$
exhibit similar behavior, although with some deviations due to
contributions from ${\bf q} \neq {\bf Q}$. In any case, deviations from
$\chi_i \sim {\cal A}_i /(T+\Theta_i)$ diminish as we consider further
neighbors or when the perturbation by the impurity is less local (or
weaker). From the theory it is clear that ${\cal A}_i$ is related to AFM
correlations $C_{0i}^0$.  Even more important, the transition 
from the regime with Kondo scale $\Theta_i \sim 0$ to a finite and fast
increasing $\Theta_i$ reflects the crossover from a non-Fermi liquid
to a  more normal Fermi-liquid regime.
 
Authors acknowledge the support of the Ministry of Education, Science
and Sport of Slovenia under grant P1-0044.

\end{document}